\documentclass[prb,aps,twocolumn,floatfix]{revtex4-1}

\usepackage{graphicx}
\usepackage{amsmath}
\usepackage{epstopdf}
\usepackage{amsbsy}
\usepackage{appendix}
\usepackage{ulem}
\usepackage{xcolor}
\usepackage{soul}
\usepackage{textcomp}
\usepackage{gensymb}

\newcommand{\beq}{\begin{equation}}
\newcommand{\eeq}{\end{equation}}
\newcommand{\bea}{\begin{eqnarray}}
\newcommand{\eea}{\end{eqnarray}}

\def\ang{\,\textrm{\AA}}
\def\VAA{\,\textrm{V/\AA}}
\def\eV{\,\textrm{eV}}
\def\meV{\,\textrm{meV}}
\def\Ry{\,\textrm{Ry}}
\def\GPa{\,\textrm{GPa}}
\def\muB{\,{\mu}_\textrm{B}}

\def\abinitio{\textit{ab initio}}

\newcommand{\rom}[1]{\lowercase\expandafter{\romannumeral #1\relax}}

\begin{document}

\title{First-principles calculation of gate-tunable ferromagnetism in magic-angle twisted bilayer graphene under pressure}

\author{Xiao Chen}
\affiliation{Department of Physics, University of Florida, Gainesville, FL 32611, USA}
\affiliation{Quantum Theory Project, University of Florida, Gainesville, FL 32611, USA} 

\author{James N Fry}
\affiliation{Department of Physics, University of Florida, Gainesville, FL 32611, USA}

\author{Shuanglong Liu}
\affiliation{Department of Physics, University of Florida, Gainesville, FL 32611, USA}
\affiliation{Quantum Theory Project, University of Florida, Gainesville, FL 32611, USA}

\author{Hai-Ping Cheng}
\email{hping@ufl.edu}
\affiliation{Department of Physics, University of Florida, Gainesville, FL 32611, USA}
\affiliation{Quantum Theory Project, University of Florida, Gainesville, FL 32611, USA}
\affiliation{Center for Molecular Magnetic Quantum Materials, University of Florida, Gainesville, FL 32611, USA}

\begin{abstract}

Magic-angle twisted bilayer graphene (MATBG) is notable as a highly tunable platform for investigating strongly correlated phenomena such as high-$T_c$ superconductivity and quantum spin liquids, due to easy control of doping level through gating and sensitive dependence of the magic angle on hydrostatic pressure. Experimental observations of correlated insulating states, unconventional superconductivity and ferromagnetism in MATBG indicate that this system exhibits rich exotic phases. In this work, using density functional theory calculations in conjunction with the effective screening medium method, we find the MATBG under pressure at a twisting angle of $2.88\degree$ and simulate how its electronic states evolve when doping level and out-of-plane electric field are gate-tuned. Our calculations show that, at doping levels between two electrons and four holes per moir\'{e} unit cell, a ferromagnetic solution with spin density localized at AA stacking sites is lower in energy than the nonmagnetic solution. The magnetic moment of this ferromagnetic state decreases with both electron and hole doping and vanishes at four electrons/holes doped per moir\'{e} unit cell. Hybridization between the flat bands at the Fermi level and the surrounding dispersive bands can take place at finite doping. Moreover, upon increasing the out-of-plane electric field at zero doping, a transition from the ferromagnetic state to the nonmagnetic one is seen. We also analyze the interlayer bonding character due to the flat bands via Wannier functions. Finally, we report trivial band topology of the flat bands in the ferromagnetic state at a certain doping level.

\end{abstract}

\maketitle{}

\section{Introduction}

The correlated insulating and superconducting phases in twisted bilayer graphene (TBG) with twisting angle $\theta\sim1.1\degree$ have attracted great interest since their initial discovery \cite{CaoYuanSC,CaoYuanInsul} owing to the similarity between their phase diagram and that of high-$T_c$ cuprates \cite{CupratePhaseDiagram,CupratePhaseDiagramSecond} and to the easy control of doping level in this two-dimensional (2D) system by gating.
This angle was theoretically predicted to be a ``magic angle'' \cite{AHMacDonald_Moire_bands_in_twisted_doublelayer_graphene}, \textit{i.e.} a twisting angle at which the Fermi velocity vanishes and the bands at the Fermi level become flat. 
The appearance of these flat bands in magic-angle twisted bilayer graphene (MATBG) is widely believed to be a signature of the strongly correlated electronic behavior in this system. Recently, in addition to the insulating and superconducting phases, MATBG was found to exhibit orbital ferromagnetism \cite{Superconductors_orbital_magnets_and_correlated_states_in_MAtBLG,Emergent_ferromagnetism_near_three-quarters_filling_in_twisted_bilayer_graphene}, which makes it a rare example of bulk magnetism that is purely orbital. This new finding further confirms the idea that MATBG is a promising platform exhibiting wide range of strong correlated, exotic phases.

Although there have been many theoretical models addressing the possible mechanism of the experimentally observed exotic behavior and predicting new phases of MATBG, the results are often sensitive to the values of parameters adopted. Therefore, it is valuable to numerically simulate the system in all its details within an \abinitio{} method that uses a minimum set of arbitrary parameters. However, the large number of atoms in a moir\'{e} unit cell of MATBG, of the order of ten thousand as a consequence of the small first magic angle, has greatly hindered thorough \abinitio{} investigation of the material since the early days.
The situation became better with the theoretical calculation of the interlayer distance, and thus pressure, dependence of the magic angles, which shows an enhancement of the first magic angle from $\sim1.1\degree$ to $\sim3\degree$ if the pressure is increased from $0$ to an experimentally accessible $30 \GPa$ \cite{MA_pressure_dependence}. Strongly correlated insulating behavior and superconductivity were also experimentally induced in TBG with twisting angle larger than 1.1 degrees, which did not exist otherwise, by tuning the interlayer distance with hydrostatic pressure \cite{Tuning_SC_in_tBLG}. The existence of flat bands at larger twisting angle under pressure renders \abinitio{} study of MATBG much more feasible, since the number of atoms per moir\'{e} unit cell decreases drastically with only minor increase of twisting angle from $1.1 \degree$ \cite{Base_structure}.  This, together with the recent experimental discovery of ferromagnetism in MATBG  \cite{Superconductors_orbital_magnets_and_correlated_states_in_MAtBLG,Emergent_ferromagnetism_near_three-quarters_filling_in_twisted_bilayer_graphene}, motivates our work of finding and thoroughly investigating the ferromagnetic ground state in gated MATBG under pressure using density functional theory (DFT) in conjunction with the effective screening medium (ESM) method, which has been proven effective in simulating gate field effects in 2D or quasi-2D solid state systems \cite{ESMmethod,RN3033,RN2180}.

In this paper, we consider MATBG with a twisting angle of $2.88\degree$ under pressure. We show the emergence of a ferromagnetic (FM) state that is lower in total energy than the nonmagnetic (NonM) state as flat bands appear, and investigate how this state evolves with electrostatic doping and out-of-plane electric field. 
Electrostatic doping results in a reduction of magnetization of the FM state for both electron and hole doping. It also leads to a significant relative shift in energy between the flat bands and surrounding dispersive bands, which can lead to crossing/hybridization between the two groups. 
An out-of-plane electric field causes a transition from the FM state to the NonM state.
In addition, we perform a Wannier analysis of the flat bands to examine interlayer bonding character and band topology. 

\section{Computational Methods}\label{sec:method}

The SIESTA density functional theory (DFT) code \cite{siesta} was used to perform self-consistent calculations of the electronic structure of gated TBG.
A double-$\zeta$ polarized atomic-like basis set and the Perdew-Zunger (PZ) exchange-correlation functional \cite{SLLRN328, SLLRN329} within the local-spin-density approximation (LSDA) were used.
Norm-conserving pseudopotentials generated via the Troullier-Martins scheme \cite{SLLRN271} were applied for carbon atoms.
First Brillouin zone (BZ) integrations were done using a uniform $6 \times 6$ Monkhorst-Pack $k$-point grid \cite{Monkhorst_Pack}. 
The real-space grid for numerical integrals has a plane-wave cutoff of $ 200 \Ry $. 
The tolerance for Hamiltonian matrix elements and in the density matrix were set to $0.001 \eV$ and $1\times10^{-4}$ respectively. 
Within the ESM method, a gate electrode is modeled by imposing a boundary condition of constant electrostatic potential. 
In both our single-gate and dual-gate simulations, TBG is separated from the gate electrode by a vacuum layer of $15 \ang{}$ thickness. 
We note that spin-orbit coupling is not included in our numerical simulations.

We obtained Maximally localized Wannier functions using the approach of Marzari and Vanderbilt \cite{SLLRN315} as implemented in the Wannier90 computational package \cite{SLLRN316}. 
A $10 \times 10$ ($6 \times 6$) uniform $k$-point mesh was sampled in  reciprocal space for the extraction of a $4$-band ($20$-band) model in the representation of Wannier functions. 
The $\Gamma$ point is not included for obtaining the $4$-band model because of a dispersive energy band crossing the flat energy bands near it. 
The $\Gamma$ point is however included when we obtain the $20$-band model. 
The convergence tolerance in the spread is set to $1 \times 10^{-6}$ and $1 \times 10^{-10} \ang^2$ for the disentanglement procedure and the Wannierization procedure respectively. 
A $500 \times 500$ uniform $k$-point mesh was applied to calculate the projected density of states based on the model Hamiltonian.

\section{Results and Discussion}\label{sec:results}

We consider TBG geometries created by rotating one layer of an AA stacking bilayer graphene with respect to the other about an axis perpendicular to the parallel flat layers which also passes through two carbon atoms \cite{Base_structure,CastroNeto_grapheneBilayerwithATwistElectronicStruct,MYChou_EffectOfEandMetalDopantsOntBLG,Magaud_LocalizationOfDiracElectronsIntBLG}. The rotation angle $\theta$ is such that the resulting atomic structure is commensurate. 
Pressure is simulated by reducing the separation between the two graphene layers.
Each of the commensurate TBGs can be labeled by a set of two co-prime integers $(n,m)$ which relate to the twisting angle as \cite{Base_structure,CastroNeto_grapheneBilayerwithATwistElectronicStruct,MYChou_EffectOfEandMetalDopantsOntBLG,Magaud_LocalizationOfDiracElectronsIntBLG} 
\begin{equation}
\cos \theta = \frac{n^{2}+4nm+m^{2}}{2(n^{2}+nm+m^{2})}.
\end{equation} 
The lattice vectors of the TBG moir\'{e} unit cell are $\mathbf{t}_{1}=n\mathbf{a}_{1}+m\mathbf{a}_{2}$ and $\mathbf{t}_{2}=-m\mathbf{a}_{1}+(n+m)\mathbf{a}_{2}$, where $\mathbf{a}_{1}$ and $\mathbf{a}_{2}$ are the lattice vectors of the primitive unit cell of monolayer graphene.

\subsection{Finding flat band condition} 

\begin{figure}[htb!]
\begin{centering}
\includegraphics[width=0.95\columnwidth]{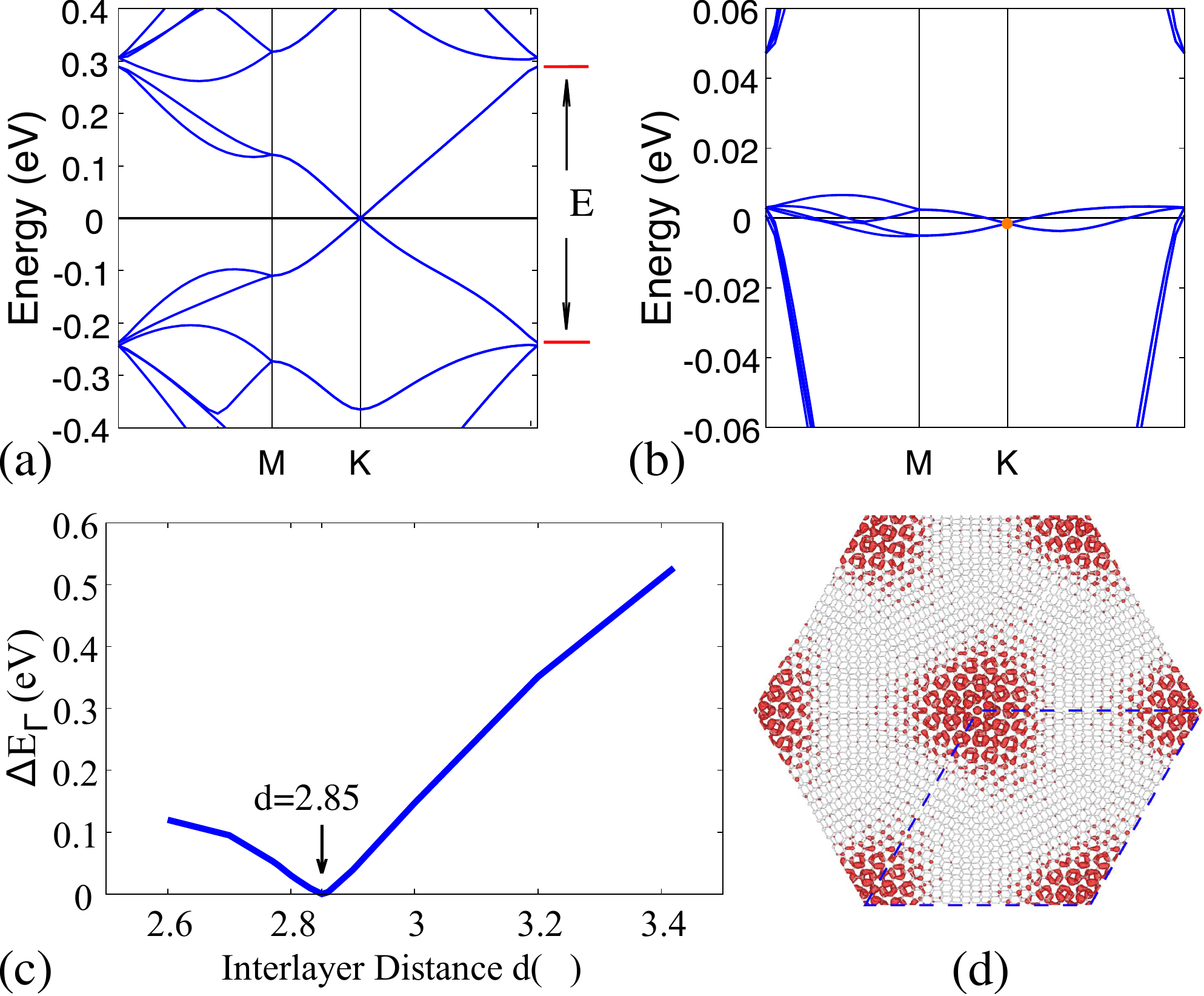}
\end{centering}
\caption{
\label{fig:flat_bands} 
(a) Band structure of $\theta=2.88\degree$ TBG with interlayer distance $d=3.42\ang$.
This corresponds to essentially zero pressure. 
(b) Band structure of $\theta=2.88\degree$ TBG with interlayer distance $d=2.85\ang$. 
(c) ${\Delta}E_{\Gamma}$ vs. interlayer distance $d$, where ${\Delta}E_{\Gamma}$ is the total width of the bands near the Fermi level as marked in the panel (a). 
(d) Modulus of a wavefunction highlighted in panel (b). 
Isosurface level: $0.006\, \textrm{Bohr}^{-3/2}$. The blue dashed line marks the boundary of a moir\'{e} unit cell.
}
\end{figure}

The TBG system we choose to study is labeled by integers (12,11) with a corresponding twisting angle of $\theta=2.88\degree$. Both for simplicity and because electronic properties of TBG are mostly smooth functions of $\theta$ but not of $(n,m)$, we will refer to the system by its twisting angle. The number of atoms per moir\'{e} unit cell is 1588, which is the local minimum among those of the commensurate TBGs that have twisting angles close to $2.88\degree$ and pushes to the limit of our computational capability. According to the work of Carr \textit{et al.} \cite{MA_pressure_dependence}, the pressure required to compress the $2.88\degree$ TBG in order for the flat bands to emerge near the Fermi level is experimentally accessible.
In the following, we decrease the interlayer distance of this system from its value at zero pressure and study the evolution of the band width at the Fermi level. All the calculations in this subsection are non-spin-polarized.

Fig.~1a shows the band structure of the $\theta=2.88\degree$ TBG with interlayer distance $d=3.42\ang$, which corresponds to essentially zero pressure.
In this case, all the bands are dispersive, and a clear Dirac point is present at $K$ at the Fermi level. 
The band that ranges from $0$ to $0.28\eV$ has 4-fold degeneracy along the $k$-point path $M\textrm{-}K\textrm{-}\Gamma$, 2-fold from spin degeneracy and 2-fold from valley degeneracy. 
This is also the case for the band that ranges from $ -0.24 \eV $ to $0$. 
Therefore, full filling of these two sets of bands corresponds to $ \pm 4$ electrons doped per moir\'{e} unit cell. 
Fig.~1b shows the band structure at the interlayer distance $d=2.85\ang$, which from our calculation corresponds to a pressure of $\sim14 \GPa$. 
At this separation, flat bands show up near the Fermi level. 
In this case, the single particle gap between the flat bands and the dispersive ones above remains, while the flat bands below become almost degenerate with the dispersive bands at the $\Gamma$ point. 
${\Delta}E_{\Gamma}$ versus the interlayer distance $d$ is plotted in Fig.~\ref{fig:flat_bands}c, where $\Delta E_{\Gamma}$ is the total width of the bands at the Fermi energy as measured at the $\Gamma$ point (see Fig.~\ref{fig:flat_bands}a); 
for instance, $\Delta E_{\Gamma}=525\meV$ at $d=3.42\ang$. 
The width ${\Delta}E_{\Gamma}$ decreases as $d$ is reduced until  $2.85\ang$, where it vanishes; 
$\Delta E_{\Gamma}$ becomes finite again for even smaller $d$, which is consistent with previous study.~\cite{MA_pressure_dependence} 
Displayed in Fig.~1d is the modulus of a flat-band wavefunction at the $K$-point of the $\theta=2.88\degree$ system with interlayer distance $d=2.85\ang$.  
The wavefunctions of flat bands exhibit strong localization at AA stacking sites, which makes sense since localized states have reduced hoping integrals from a tight-binding point of view, and this usually indicates band flatness. Here we have found a MATBG system of $\theta=2.88\degree$ and $d=2.85\ang$ under pressure that will be our focus in the remaining part of the paper.

\subsection{Evolution of ferro- and non-magnetic solutions under  electrostatic doping}

Next, we investigate single-gate field effects on the $\theta=2.88\degree$ TBG with interlayer distance $d=2.85\ang$, which possesses flat bands around the Fermi energy. 
A single-gate setup allows electrostatic charge doping of the TBG.

\begin{figure}[htb!]
\begin{centering}
\includegraphics[width=0.95
\columnwidth]{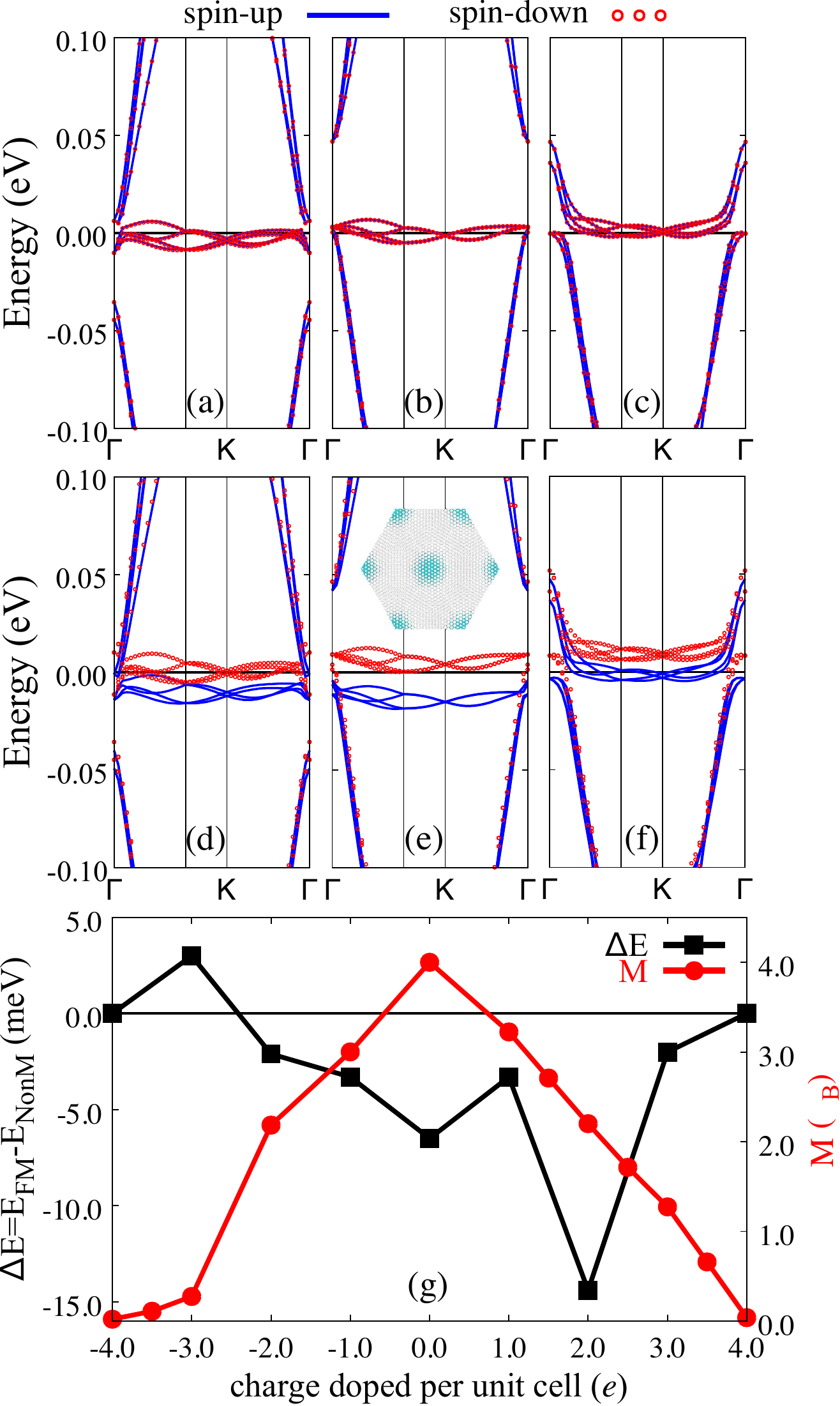}
\end{centering}
\caption{
\label{fig:M_v_n} 
(a)--(c) [(d)--(f)] Band structures of the nonmagnetic [ferromagnetic] solution for TBG with $\theta=2.88\degree$ and $d=2.85\ang$ under doping levels of two electrons, charge neutral, and two holes per moir\'{e} unit cell, respectively. 
The inset of panel (e) shows the spin density of the ferromagnetic solution of the neutral system, which shows strong localization at AA stacking sites. 
The isosurface level is $3\times10^{-4}\, \ e^- \, \textrm{Bohr}^{-3}$. 
(g) Energy difference between the ferromagnetic and the nonmagnetic solutions and magnetization per moir\'{e} unit cell of the ferromagnetic solution at each doping level.
$e$ is the unit charge. 
A positive (negative) doping level means hole (electron) doping. 
}
\end{figure}

Figs.~\ref{fig:M_v_n}a--\ref{fig:M_v_n}c display the band structures of the NonM solution of the system under doping levels of two electrons, charge neutral and two holes per moir\'{e} unit cell, respectively. 
Upon finite doping, the valley degeneracy of the flat bands along the $k$-point path $M\textrm{-}K\textrm{-}\Gamma$ is lifted due to gate-induced electrostatic potential difference between the two graphene layers. 
At most doping levels between 4 electrons per moir\'{e} unit cell and 4 holes per moir\'{e} unit cell, both the flat bands and the dispersive bands around the Fermi energy are doped and thus partially occupied. 
The shift of the dispersive bands in energy relative to $E_F$ due to doping is much larger than that of the flat bands. 
As a result, the dispersive bands cross the flat bands near the $\Gamma$ point, where a single particle gap opens due to their hybridization.

Figs.~\ref{fig:M_v_n}d--\ref{fig:M_v_n}f show the band structures of the FM solution of the system under the same doping levels as Figs.~\ref{fig:M_v_n}a--\ref{fig:M_v_n}c, respectively. 
Here both the valley and spin degeneracy of the flat bands are broken at finite doping levels, resulting in an unbalanced occupation between the spin-up and spin-down channels and thus a net magnetic moment.
The inset of Fig.~\ref{fig:M_v_n}e is the spin density in real space of the FM solution of the neutral system, displaying strong localization at AA stacking sites. 
This is because spin splitting mainly occurs in the flat bands, which are by themselves localized at these sites. 
Thus the spin density, defined as the difference between spin-up and spin-down electron densities, can only be localized also at AA stacking sites. This localization persists through all doping ranges from four electrons to four holes doped per moir\'{e} unit cell.

Finally, in Fig.~\ref{fig:M_v_n}g, we show the energy difference between the FM and NonM solutions and the magnetization per moir\'{e} unit cell of the FM solution at different doping levels. 
The total energy of the FM solution is lower than that of the corresponding NonM one at doping levels ranging from 2 electrons to 4 holes doped per moir\'{e} unit cell.
In the case of zero doping (see Figs.~\ref{fig:M_v_n}b and \ref{fig:M_v_n}e), this can be understood by examining the band energy: The spin-up (spin-down) flat bands are lowered down (lifted up) in energy and become fully occupied (nearly empty) after spin splitting. 
The FM state is stabilized the most at the doping level of 2 holes per moir\'{e} unit cell.
However, the NonM solution has lower energy at doping levels between three and four electrons per moir\'{e} unit cell. The magnetization near zero doping is the highest, and it decreases monotonically as the doping level increases in both the directions of electron and hole doping. 
It vanishes at both four electrons and four holes doped per moir\'{e} unit cell. 

It is worth mentioning here that we have also calculated  $\theta=2.88\degree$ TBGs with interlayer distances $d=2.83\ang$ and $2.87\ang$ at doping levels of $\pm2$ charges per moir\'{e} unit cell. The FM solution is not found for the $d=2.87\ang$ systems, while in the $d=2.83\ang$ cases, an FM solution with magnetization per moir\'{e} unit cell $M$ of only $0.5 \muB $ ($0.3 \muB $) exists at the doping level of 2 holes (electrons) per moir\'{e} unit cell. Comparing these to the magnetizations of the $d=2.85\ang$ systems at the same 2 hole and 2 electron doping, which both have $M=2.2\muB $, we arrive at the conclusion that the FM solution only exists in a narrow range of interlayer distance around the flat-band condition.

Lopez-Bezanilla studied substitutional chemical doping of $\theta = 5.09 \degree$ TBG under pressure and found that $N$-doping can turn ferromagnetic order into antiferromagnetic order \cite{SLLRN347}. 
Compared with chemical doping, electrostatic doping provides continuous and reversible control over the doping level without destroying intrinsic properties of the system under study. 
However, limited by the large size of our $\theta=2.88\degree$ system, we included only one moir\'{e} unit cell in our simulations, and thus it remains an open question whether antiferromagnetic order occurs upon electrostatic doping.

\subsection{Hybridization of flat and dispersive bands}

\begin{figure}[htp]
\begin{centering}
\includegraphics[width=0.9\columnwidth]{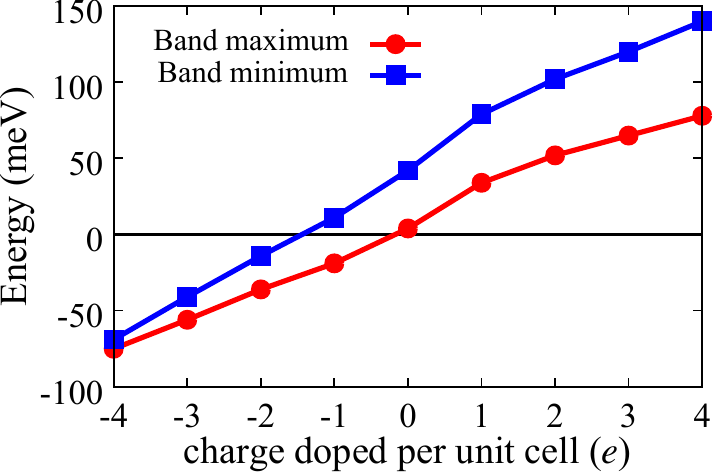}
\end{centering}
\caption{\label{fig:band_edge}
Circles [squares] show the energy relative to the Fermi level of the band maximum [minimum] of the dispersive bands below [above] the flat bands. Here the data is for the nonmagnetic solution.
}
\end{figure}

Atomic structure relaxation in \abinitio{} calculations is known to open and widen a single particle gap below as well as above the flat bands \cite{MA_pressure_dependence,RelaxationGap_theory}.
This gap in the case of zero pressure MATBG is known to be less than $\sim50\meV$ both from \abinitio{} calculations \cite{MA_pressure_dependence,RelaxationGap_theory} and from experiments~\cite{CaoYuanInsul,CaoYuanSingleParticleGap}, and its size has been shown to decrease if the twisting angle is increased from $1.1\degree$.
However, the relative shift in energy between the dispersive and flat bands due to charge doping varies at a much larger scale in our calculation of the $\theta=2.88\degree$ and $d=2.85\ang$ system.
This can seen from how the energy, relative to the Fermi level, of the band maximum (minimum) of the dispersive bands below (above) the flat bands evolves with the doping level, as shown in Fig.~\ref{fig:band_edge} for the NonM solutions.
Since the energy of the flat bands is not sensitive to the doping level within the range of $\pm4$ charges doped per moir\'{e} unit cell (see Figs.~\ref{fig:M_v_n}a--\ref{fig:M_v_n}c), the band edges plotted here are a good measure of the hybridization between the dispersive bands and the flat bands. 
The maximum (minimum) of the dispersive bands below (above) the flat bands moves up (down) in energy with respect to the flat bands by $\sim 150$ ($\sim 200$) meV, when the doping level varies from 4 electrons (holes) to 4 holes (electrons) doped per moir\'{e} unit cell.
Although a full atomic structure relaxation under pressure is beyond the scope of this work, the large size of the relative shift in energy between the dispersive and flat bands shown here indicates that the common assumption that the flat bands are perfectly isolated from the surrounding dispersive ones may not hold true for MATBG at large twisting angles at certain finite doping levels.

\subsection{Effect of out-of-plane electric field}

\begin{figure}[htb!]
\begin{centering}
\includegraphics[width=0.95\columnwidth]{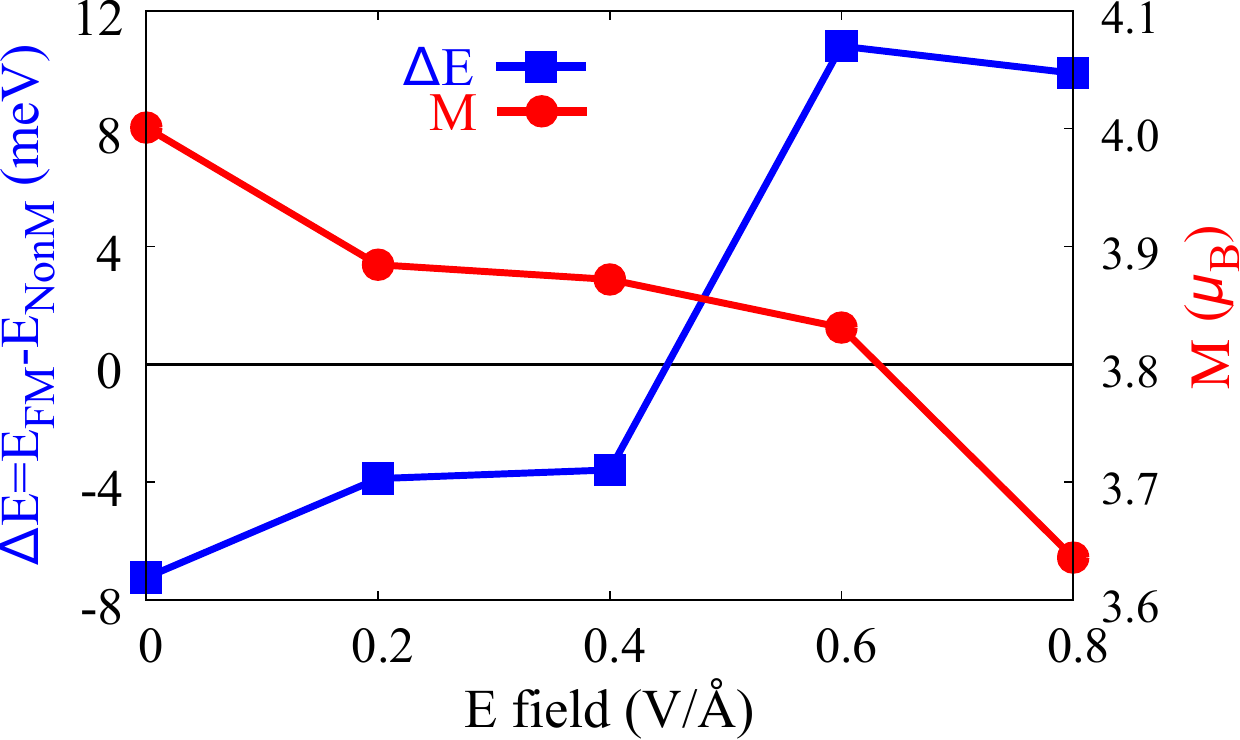}
\end{centering}
\caption{
\label{fig:Ex_effects_M_and_E} 
Circles: Magnetization per moir\'{e} unit cell $M$ of the ferromagnetic solution as a function of the strength of the out-of-plane electric field. 
Squares: Energy difference between the ferromagnetic state and the nonmagnetic state. }
\end{figure}

A dual-gate setup permits an out-of-plane electric field between two gate electrodes as well as charge doping. 
In this subsection, we examine the effect of out-of-plane electric field on the charge neutral $\theta=2.88\degree$ TBG with interlayer distance $d=2.85\ang$. 
Since our dual-gate setup is symmetric, a negative out-of-plane electric field is equivalent to an oppositely directed electric field with the same magnitude. 
As such, it is sufficient to consider positive electric fields applied to the TBG. 
Fig.~\ref{fig:Ex_effects_M_and_E} shows the magnetization per moir\'{e} unit cell $M$ of the FM solution versus the strength of the out-of-plane electric field. 
$M$ decreases with the electric field from $4.0 \muB$ at zero field. 
This decrease is slow and only $0.17\muB$ ($4.2\%$) before $0.6 \VAA$, which is already a large field in experiments.
Fig.~\ref{fig:Ex_effects_M_and_E} also shows the energy difference between the FM state and the NonM state as a function of the electric field. 
Notably, a phase transition from the FM state to the nonmagnetic state takes place between $0.4\VAA$ and $0.6\VAA$.

\begin{figure}[htp]
\begin{centering}
\includegraphics[width=0.95\columnwidth]{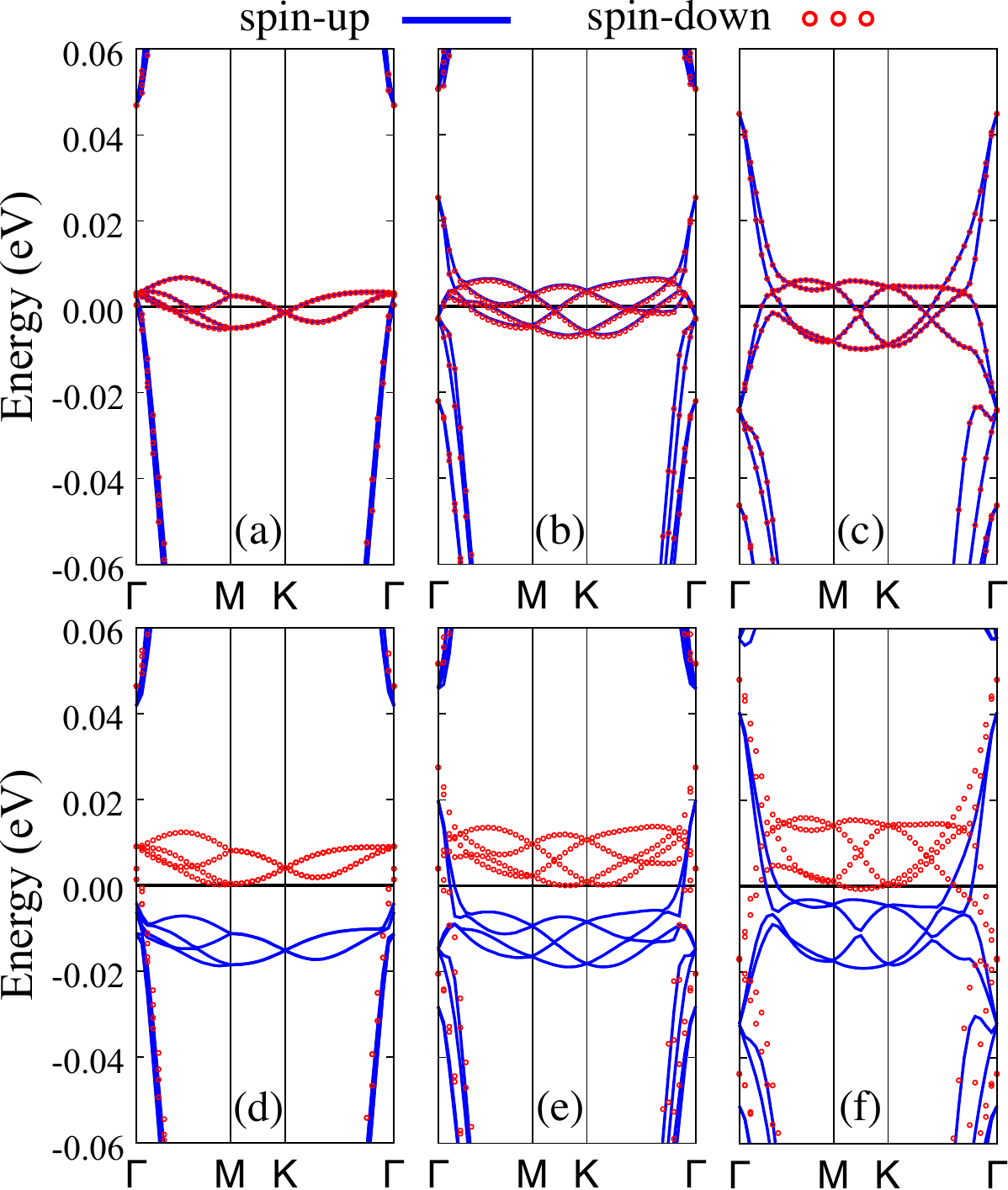}
\end{centering}
\caption{
\label{fig:Ex_effect_bands} 
(a)--(c) The band structure of the nonmagnetic solution at out-of-plane electric fields of 0, 0.4 and $ 0.8 \VAA$ respectively.
(d)--(f) The band structure of the ferromagnetic solution at out-of-plane electric fields of 0, 0.4 and $ 0.8\VAA$ respectively.
}
\end{figure}

The reason for such a phase transition can be understood in the band picture in terms of a competition between the spin splitting energy of the FM state and the energy splitting in the valley degree of freedom
Fig.~\ref{fig:Ex_effect_bands}d shows the band structure of the FM solution at zero electric field, where the spin-up flat bands are lowered in energy to be fully occupied while the spin-down flat bands become nearly empty. 
After such a spin splitting, the occupied flat-band states have lower energy than those of the nonmagnetic solution at zero electric field as shown in Fig.~\ref{fig:Ex_effect_bands}a. 
This is the major reason for the FM state being energetically favorable at small out-of-plane electric fields. 
As the strength of out-of-plane electric field increases, the valley splitting of flat bands (along the $k$-point path $M\textrm{-}K\textrm{-}\Gamma$) of the nonmagnetic solution increases monotonically, which can be seen from Figs.~\ref{fig:Ex_effect_bands}a--c for the out-of-plane electric fields of 0, 0.4 and $ 0.8 \VAA$ respectively. 
Consequently, the energies relative to the Fermi level of the occupied states are mostly lowered, which contributes to a decrease in the total energy of the NonM state $E_\textrm{NonM}$. 
In contrast, the spin splitting of flat bands of the FM solution is insensitive to the out-of-plane electric field, although the valley splitting for each spin channel is comparable to that of the corresponding nonmagnetic solution (see Figs.~\ref{fig:Ex_effect_bands}d--f). 
It is also noteworthy that the spin splitting of flat bands of the FM solution is not sensitive to either the valley band index or to the crystal momentum. 
Overall, the energy of the FM state $E_\textrm{FM}$ does not vary as much as the NonM state with the out-of-plane electric field, as long as the spin-up (spin-down) flat bands remain fully occupied (empty). 
Therefore, the energy difference $E_\textrm{FM} - E_\textrm{NonM}$ increases with the electric field before $0.4 \VAA$. 
At an out-of-plane electric field $\geq 0.4 \VAA$, the valley splitting is so large that the otherwise-empty spin-down flat bands become partially occupied close to the Fermi level along $M\textrm{-}K$ cuts in the BZ, raising the energy of the FM state. 
Consequently, the energy difference $E_\textrm{FM} - E_\textrm{NonM}$ increases greatly between $0.4 \VAA{}$ and $0.6 \VAA{}$, accompanied by an FM-to-NonM phase transition.

\subsection{Wannier downfolding}

\begin{figure}[htb!]
\centering
\includegraphics[width=0.95\columnwidth]{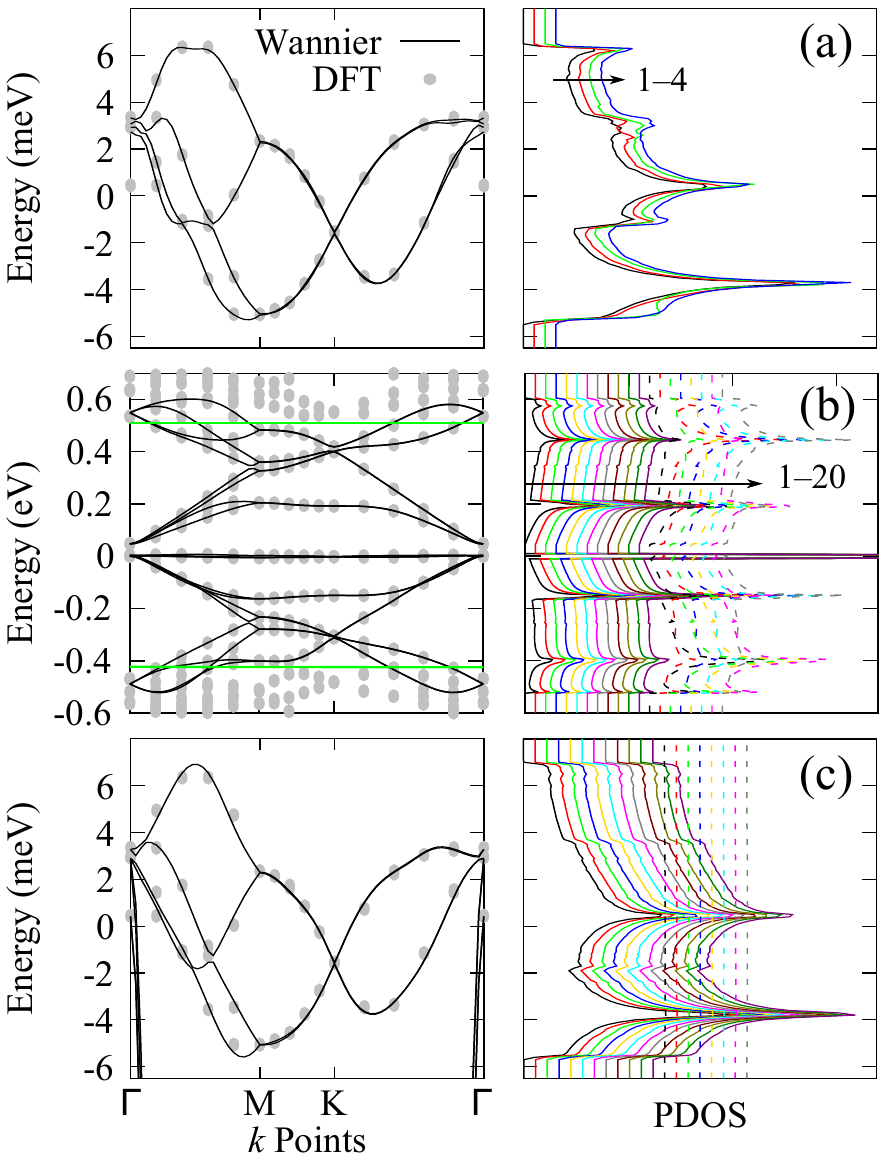}
\caption{
\label{fig:banddos} 
Band structure (left panels) of the model Hamiltonian based on Wannier functions and density of states (right panels) projected onto each Wannier orbital. 
The eigenvalues obtained by DFT are also shown for comparison. 
For panels (a) [(b) and (c)], 4 [20] energy bands around the Fermi energy are transformed into Wannier functions.
Panel (c) is the same as panel (b) except for a much smaller energy range. 
The green lines in panel (b) mark the frozen energy window used for Wannierization. 
The PDOS curves are shifted so that they can be distinguished. 
The value of the PDOS for all Wannier functions is zero at $-0.6 \eV $ for panel (b) or at $-6 \meV $ for panels (a) and (c). 
}
\end{figure}

In this subsection, we examine the bonding character in TBG under pressure via Wannier functions, which have been applied to obtain accurate tight-binding models for both monolayer graphene~\cite{SLLRN327} and AB stacked bilayer graphene~\cite{SLLRN326}. 
We performed the  Wannier analysis for non-spin-polarized TBG since spin-polarization makes little difference in the shape of the flat energy bands. 
The left panels of Fig.~\ref{fig:banddos}a (Figs.~\ref{fig:banddos}b and \ref{fig:banddos}c) shows both the DFT band structure and the interpolated band structure based on a 4-band (20-band) model Hamiltonian in the Wannier function representation. 
As can be seen from the figures, they match well for both the 4-band model and the 20-band model. 
The right panels of Fig.~\ref{fig:banddos} show the projected density of states (PDOS) projected onto each Wannier function. 
In the 4-band model, all four Wannier functions contribute almost equally to each flat band. 
In the 20-band model, the twenty Wannier functions are of two types; 
the first (type 1) consists of Wannier functions of indices ranging from 1 to 12; and the second (type 2) consists of those of indices ranging from 13 to 20. 
Fig.~\ref{fig:banddos}b shows that both Wannier function types contribute significantly to the dispersive bands, with more contribution from type 2 than that from type 1. 
Fig.~\ref{fig:banddos}c shows that only Wannier functions of type 1 contribute to the flat bands; the contribution from any Wannier function of type 2 is essentially zero.

\begin{figure}[htp!]
\centering
\includegraphics[width=0.85\columnwidth]{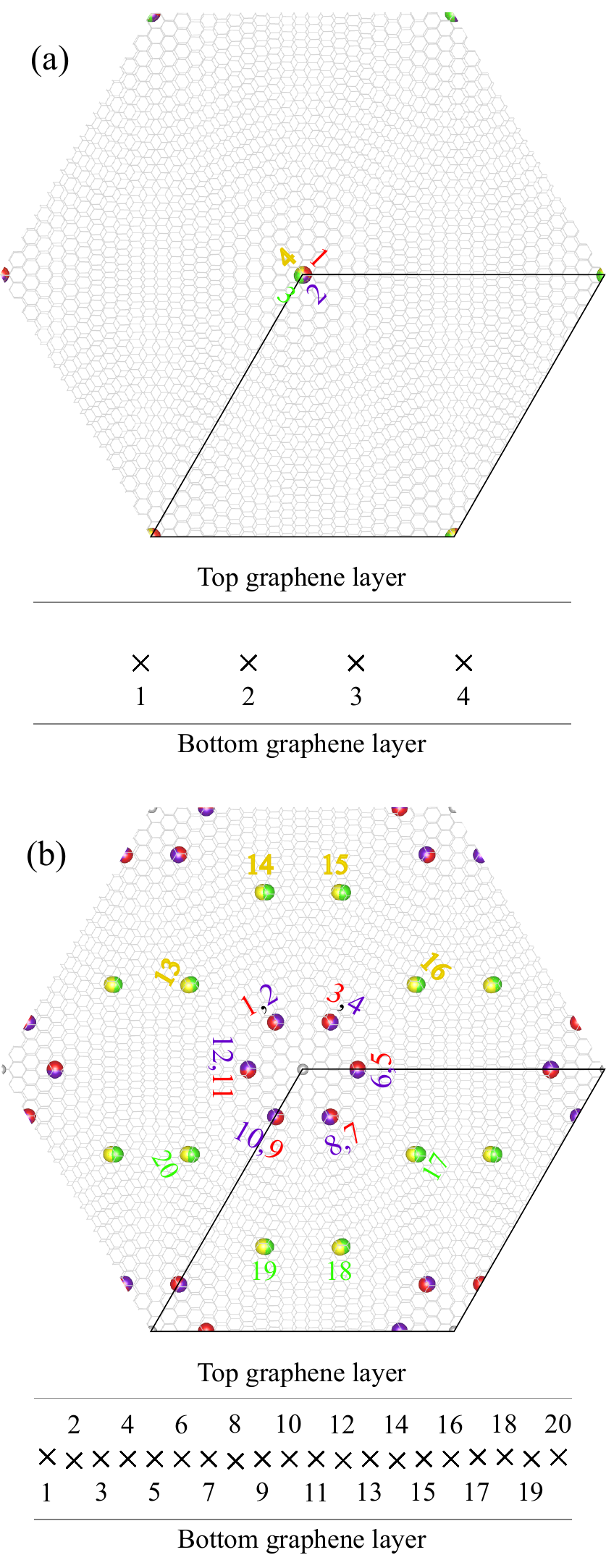}
\caption{
\label{fig:wcc} 
Positions of Wannier charge centers (WCCs) for each Wannier function. Panel (a) is for the 4-band model and panel (b) for the 20-band model. The top (side) view is shown in the upper (lower) panels. Different colors in the top view are used to differentiate multiple WCCs at almost the same position. Only WCCs belonging to the same moir\'{e} unit cell (dark solid line) are labeled by their indices. In the side view, only the vertical position is meaningful; the horizontal position is set manually. 
}
\end{figure}

Fig.~\ref{fig:wcc}a (Fig. \ref{fig:wcc}b) shows the Wannier charge centers (WCC) for the 4-band (20-band) model. 
In the 4-band model, from a top view all four WCCs are located almost at AA stacking sites. 
The flat-band states are localized at AA sites, which is consistent with previous DFT calculations~\cite{Base_structure} and STM measurements~\cite{SLLRN318, SLLRN319, SLLRN320}. 
From a side view, all four WCCs are in the middle of the two graphene layers. 
This indicates that each flat band mediates certain bonding between the two graphene layers (via the AA site). 
In the 20-band model, from a side view all twenty WCCs are in the middle of the two graphene layers as well. 
However, the in-plane positions of the WCCs are quite different from those in the 4-band model: 
type 1 (type 2) WCCs are closer to AA sites (AB sites). 
Specifically, each type 1 (type 2) WCC is about $0.181 \, a_0$ 
($0.131 \,  a_0$) away from an AA (AB) site, 
where, $a_0$ is the period of the moir\'{e} pattern. 
That the WCCs are away from AA sites in the 20-band model is likely due to the flat bands being mixed with the dispersive bands. 
We infer that the dispersive bands mediate bonding between the two graphene layers via sites other than AA sites. 
 
% \onecolumngrid

% \twocolumngrid

\begin{figure}[htp!]
\centering
\includegraphics[width=0.85\columnwidth]{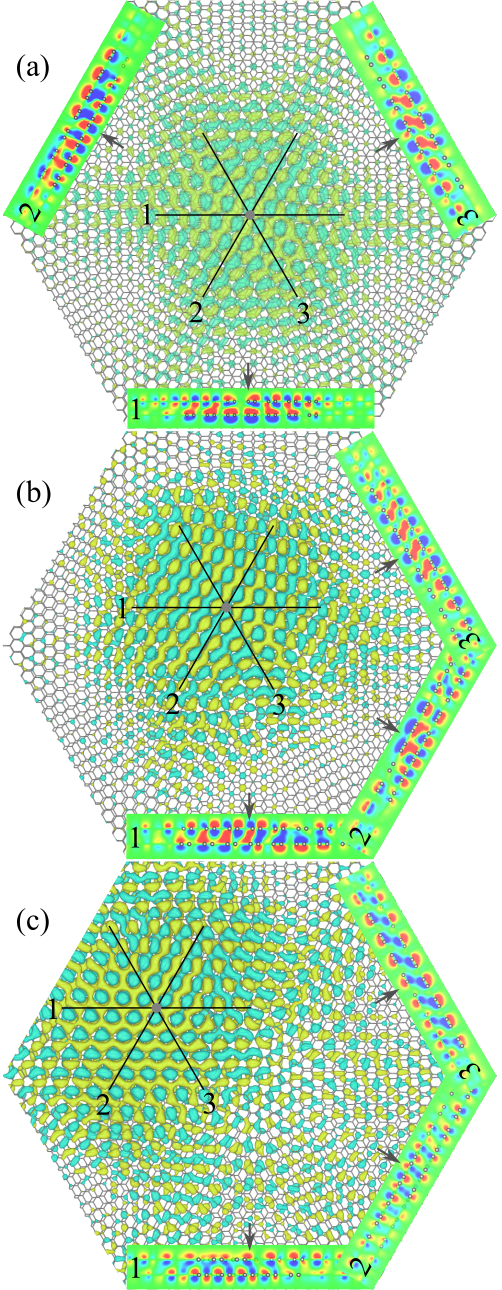}
\caption{
\label{fig:wanf} 
Isosurface and cross section of the real part of three typical Wannier functions. (a) The 1st Wannier function of the 4-band model. (b) [(c)] The 1st [13th] Wannier function of the 20-band model. The isosurface level is set to $+0.002$ (yellow) and $-0.002$ (cyan) $\textrm{\AA}^{-3/2}$.  It is mapped to color in three cross sections passing the Wannier charge center (gray dot or black arrow) for each Wannier function. Color scheme for the color map: $\leq 0.01$ (blue), $0$ (green), and $\geq 0.01$ (red), in units of $\textrm{\AA}^{-3/2}$.
}
\end{figure}

Fig.~\ref{fig:wanf}a shows the real part of the first Wannier function for the 4-band model. 
The isosurface level is about $2.3\%$ of the maximum value of the real part in the whole space. 
The spread of this Wannier function is about $18.1\%$ of the area of the moir\'{e} unit cell. 
Note that the spread $\Omega$ is defined as 
\begin{equation}
    \Omega = \left\langle w_{n 0} \left|r^{2}\right| w_{n 0} \right\rangle-\left|\left\langle w_{n 0}|\mathbf{r}| w_{n 0})\right\rangle\right|^{2},
\end{equation} 
where $w_{n 0}$ denotes the $n\textrm{th}$ Wannier function that belongs to the home unit cell. 
Fig.~\ref{fig:wanf}a also shows three cross sections that pass the WCC as indicated by the gray dot over the isosurface or by the black arrow near each cross section.
It can be seen from the first (second) cross section that $\pi$-like bonds between neighboring carbon atoms occur near the WCC in the bottom (top) graphene layer.  
As such, interlayer $\pi$-$\pi$ hybridization seems to be present near the WCC. 
If we examine a position that is far away from the WCC, we find that the intralayer $\pi$ bonding between neighboring carbon atoms becomes weaker, meanwhile the signature of individual $p_z$ orbital becomes stronger. 
Figs.~\ref{fig:wanf}b and \ref{fig:wanf}c show the first Wannier function (type 1) and the thirteenth Wannier function (type 2) respectively for the 20-band model. 
The spread of the first (thirteenth) Wannier function is about $13.2\%$ ($27.4\%$) of the area of the moir\'{e} unit cell. 
It is reasonable that the first Wannier function has a much smaller spread than the thirteenth Wannier function, 
since the former originates partially from the flat bands while the latter originates mostly from the dispersive bands. 
The cross sections in Fig.~\ref{fig:wanf}b show that some $\pi$-like orbital of the top graphene layer hybridizes with some $p_z$-like orbital of the bottom graphene layer near the WCC. 
Even more pronounced $p_z$-like orbital character can be observed in Fig.~\ref{fig:wanf}c. 
We may understand this as follows: the 20-band model involves dispersive energy bands which possess mainly $p_z$ orbital character, as in the case of monolayer graphene~\cite{SLLRN327}. 
It is still true for the 20-band model that $p_z$ orbital character becomes obvious at positions far away from the WCC.

\begin{figure}[htb!]
\centering
\includegraphics[width=0.95\columnwidth]{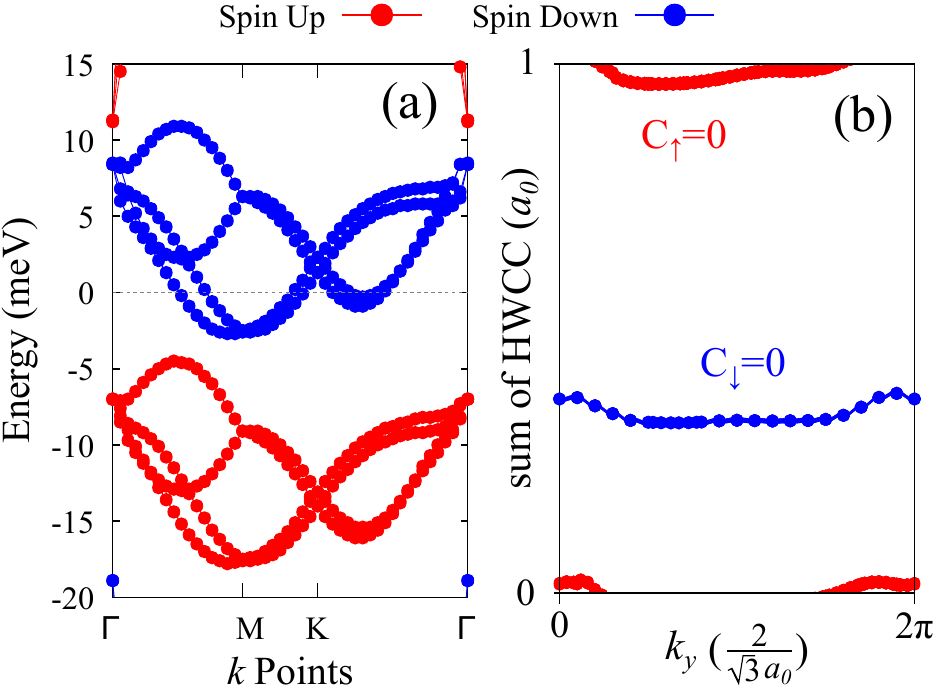}
\caption{
\label{fig:chern} 
(a) Band structure of twisted bilayer graphene doped with one electron per moir\'{e} unit cell under pressure. 
The Fermi energy is set to zero, as indicated by the dashed gray line. 
(b) Sum of the hybrid Wannier charge centers (HWCC) for the flat bands around the Fermi energy. 
$a_0$ is the period of the moir\'{e} pattern. 
$2\pi \, ({2}/{\sqrt{3}a_0}) $ is the length of the basis vector $\mathbf{b}_y$ in reciprocal space. 
}
\end{figure}

Lu \textit{et al.}~\cite{Superconductors_orbital_magnets_and_correlated_states_in_MAtBLG} reported states with non-zero Chern numbers in MATBG with one electron or one hole doped per moir\'{e} unit cell and under certain magnetic field.
Electrically tunable Chern number was also found in an ABC-trilayer graphene/hexagonal boron nitride moir\'{e} superlattice~\cite{SLLRN325}.
Inspired by these studies, we examine the band topology of the $\theta=2.88\degree$ TBG under pressure, namely with interlayer distance $d=2.85\ang$. 
The characteristic topological invariant Chern number is defined by~\cite{SLLRN321, SLLRN322} 
\begin{equation}
    C=\frac{1}{2 \pi} \int_{S} \sum_{n \in B} \nabla_{\mathbf{k}} \times i \left\langle u_{n \mathbf{k}}\left|\nabla_{\mathbf{k}}\right| u_{n \mathbf{k}}\right\rangle \cdot \mathrm{d} \mathbf{S},
\end{equation}
where $n$ is the band index, $B$ is an isolated set of energy bands, $u_{n \mathbf{k}}$ is the periodic part of a Bloch wave, and $S$ is a closed orientable two-dimensional surface (the whole Brillouin zone in our case). 
For the sake of the flat bands being isolated from the rest energy bands, we chose the case of one electron doping per moir\'{e} unit cell (see Fig.~\ref{fig:chern}a). 
It is noteworthy that the TBG exhibits half-metallicity in this case. 
In practice, the Chern number can be found by examining how the sum of hybrid Wannier charge centers (HWCC) changes with crystal momentum~\cite{SLLRN323}. 
Fig.~\ref{fig:chern}b shows the sum of HWCCs versus the crystal momentum $k_y$, which was calculated using the Z2Pack package based on a 20-band model Hamiltonian in the representation of Wannier functions~\cite{SLLRN324}.
Since the winding number of the curve for each spin in Fig.~\ref{fig:chern}b is zero, the Chern numbers $C_\uparrow$ (for spin up) and $C_\downarrow$ (for spin down) are both zero. 
As such, the flat bands in each spin channel of the ($\theta=2.88\degree$, $d=2.85\ang$) TBG with one electron doped per moir\'{e} unit cell have trivial band topology according to our calculations. 
It is unknown whether non-trivial band topology occurs for the TBG at a different doping level. 
Previous studies suggest that non-trivial band topology may occur in TBG if strong correlation is taken into account~\cite{Superconductors_orbital_magnets_and_correlated_states_in_MAtBLG, SLLRN335}.

\section{Conclusion}\label{sec:Conclusion}

In the present work, we have studied effects of electrostatic doping and transverse electric field on the nonmagnetic and ferromagnetic states in magic-angle twisted bilayer graphene of $2.88\degree$ under pressure using density functional theory in conjunction with the ESM method.
We have identified: 
(\rom{1}) that the ferromagnetic state is lower in total energy than the nonmagnetic one when the doping level is between two electrons and four holes per moir\'{e} unit cell. 
(\rom{2}) monotonic suppression of magnetic moment of the ferromagnetic state upon both electron and hole doping from charge neutrality.
(\rom{3}) hybridization between the flat bands and the surrounding dispersive ones when the system is electrostatically doped, indicating that the picture of perfectly isolated flat bands may not hold true for MATBGs under pressure with large twisting angles. 
(\rom{4}) a phase transition from the ferromagnetic to the nonmagnetic state with a transverse external electric field.
Our Wannier analysis reveals interlayer $\pi$-$\pi$ hybridization near the Wannier charge center for the flat bands. 
The set of four flat bands in each spin channel of the $2.88\degree$ magic angle twisted bilayer graphene under pressure with one electron doping per moir\'{e} unit cell has trivial band topology.

\begin{acknowledgments}

This work was supported by the US Department of Energy (DOE), 
Office of Basic Energy Sciences (BES), under Contract No. DE-FG02-02ER45995. 
Computations were done using the utilities of the National Energy Research Scientific Computing Center 
and University of Florida Research Computing.

\end{acknowledgments}

\bibliographystyle{apsrev4-2}
\bibliography{refs}

\end{document}